\RequirePackage{fix-cm}

\documentclass[natbib,smallextended]{svjour3}                     

\smartqed  
\usepackage{graphicx}
\usepackage{url,lineno}
\usepackage{mathptmx}      
\usepackage{amsmath}
\usepackage{latexsym}
\usepackage{color}
\usepackage[T1]{fontenc}

\usepackage{mathptmx}      
\usepackage{latexsym}
\begin{document}
\title{Fractional Mellin Transform - A possible application in CFT 
}

\titlerunning{Fractional Mellin Transform}        

\author{R. A. Treumann \& W. Baumjohann \\ 
  }

\authorrunning{R. A. Treumann \& W. Baumjohann} 

\institute{R. A. Treumann \at International Space Science Institute, Bern, Switzerland \\ \email{artt@issibern.ch}             \and
           W. Baumjohann \at
              Space Research Institute, Austrian Academy of Sciences, Graz Austria\\ \email{wolfgang.baumjohann@oeaw.ac.at} \\
}

\date{Received: date / Accepted: date}

\maketitle

\begin{abstract}
We propose a fractional variant of Mellin's transform which may find an application in the Conformal Field Theory. Its advantage is the presence of an arbitrary parameter which may substantially simplify calculations and help adjusting convergence.
\end{abstract}

\section{Introduction}
\label{intro}
A recently proposed fractional Laplace transform \citep{treumann2014} lacks so far any application. A close relative of Laplace transforms, obtained from the bi-directional Laplace transform, is Mellins's transform of a function $f(x)$  which is defined as 
\begin{equation}\label{eq-one}
\mathcal{M}\!\left\{ f(x)\right\} \equiv\phi(s)=\int\limits_0^\infty x^{s-1}f(x)\mathrm{d}x, \quad  \mathcal{M}^{-1}\!\left\{\phi(s)\right\}\equiv f(x)=\frac{1}{2\pi i}\!\!\!\int\limits_{c-i\infty}^{c+i\infty}x^{-s}\phi(s)\mathrm{d}s
\end{equation}
Like in the Laplace transform, integration in the inverse $ \mathcal{M}^{-1}\{\phi\} $ is along a parallel to the imaginary axis at $\emph{\textsf{Re}}\,\{s\}=c>0$ to the right of all poles. 

Mellin's transform is a power law transform with restricted convergence properties imposed on the function $f(x)$. Its inverse is of particular interest when applied to expansions in inverse powers of $x$. This property of the inverse Mellin transform was recently exploited \citep{mack2009,mack2010} in conformal field theory (CFT) and dual Anti-de Sitter cosmology (AdS/CFT). Mellin amplitudes $\phi(s)$ are defined via the first of the above two integrals. Here we propose its fractional generalisation which may be useful in application to the CFT/AdS problematics for reasons noted below.

\section*{Definition}  
Let us introduce the variable $z=\log x$ in the inverse Mellin transform. Then $x^{-s}\equiv [\exp(-z)]^s$. Replacing $\exp(-z)\to (1+z/\kappa)^{-\kappa}$ with $0<\kappa\in\textsf{R}$ enables to rewrite the inverse Mellin transform into the equivalent (inverse) fractional Mellin-$\kappa$ transform
 \begin{equation}
\mathcal{M}^{-1}_\kappa\!\left\{\phi_\kappa(s)\right\}\equiv F(z)=\frac{1}{2\pi i}\!\int\limits_{c-i\infty}^{c+i\infty}\frac{\mathrm{d}s\,\phi_\kappa(s)}{(1+sz/\kappa)^{\kappa}}
\end{equation}
Conversely, with $\mathrm{d}x=x\mathrm{d}z$ the forward Mellin transform, the definition of the Mellin-$\kappa$ amplitude $\phi_\kappa(s)$, becomes
\begin{equation}
\mathcal{M}_\kappa\! \left\{F(z)\right\} \equiv\phi_\kappa(s) = \int\limits_0^\infty(1+sz/\kappa)^{\kappa}\mathrm{d}z F(z)
\end{equation}
the function $F(z)$ being subject to the convergence requirements on $f(x)$. In the limit $\kappa\to\infty$ this replacement becomes exact, as is verified applying l'Hospitale's rule. In this way the ordinary results can easily be gained if necessary after performing the calculation with $\mathcal{M}^{-1}_\kappa$. As a generalisation of the Mellin transform to arbitrary $\kappa<\infty$ it adds some useful freedom.  

One notes that the integrand of the inverse Mellin-$\kappa$ transform is regular for all $z\neq-\kappa/s$, corresponding to $x\neq \exp(-\kappa/s)$, where the integral becomes singular. For non-integer $\kappa$ it will, however, develop branch points. If necessary, $\kappa\in\textsf{C}$ can also be chosen a complex number of positive real part. 

The definition of the inverse Mellin-$\kappa$ transform is not unique, however, for the reason that $[\exp{-z}]^s\equiv\exp(-sz)$, the form we used in the above. When referring to the first version, the Mellin-$\kappa$ transform becomes
\begin{equation}
\mathcal{M}^{-1}_{1\kappa}\!\left\{\phi_\kappa(s)\right\}\equiv F(z)=\frac{1}{2\pi i}\!\int\limits_{c-i\infty}^{c+i\infty}\frac{\mathrm{d}s\,\phi_\kappa(s)}{(1+z/\kappa)^{s\kappa}}
\end{equation}
with variable $s$ appearing as a factor on $\kappa$ in the power. Since $s$ is a complex continuous variable, this second form is more complicated. Which version is more useful will depend on the application. Presumably and, in particular, as long as $\kappa$ is taken real, the former version $\mathcal{M}_\kappa$ will be the simpler one. Here the complex integration variable $s$ appears as a variable, not as a power. For this reason, application of functional calculus to the discussion of poles and branch cuts should become easier, in many cases simplifying the calculation.  

\section*{Application}
The application we have in mind is to conformal field theory (CFT) and its dual Anti-de Sitter cosmology \citep{malda1998}. In CFT the $n$-point correlation function in the operator product expansion (OPE) for any dimension $D\geq 2$ is a function of $\frac{1}{2}n(n-3)$ independent anharmonic ratios $\omega_\ell$ of distances $\mathbf{x}^2_{ij}=[\mathbf{x}_i-\mathbf{x}_j]^2, i<j\in\textsf{N}$ \citep{oster1973,oster1975}. For example \citep{mack2010}, the euclidean Green's function in $n=4$ is given by
\begin{equation}
G_{i_4\dots i_1}(\mathbf{x}_{i_1}\dots \mathbf{x}_{i_4})=\prod_{i<j}\left(\mathbf{x}_{ij}^2\right)^{-\xi_{ij}^0}F_{i_4\dots i_1}(\omega_1,\omega_2,\omega_3)
\end{equation}
where the exponents depend on the dimensions $d_j$ of the underlying fields $\psi^{i_j}$. They obey symmetry $\xi_{ij}^0=\xi_{ji}^0$, sum up to $\sum_j\xi_{ij}^0=d_i$, and the functions $F$ satisfy some simple permutation symmetries \citep{mack2009}. This form of the CFT Green's function is suggestive for application of the Mellin transform. 

Exactly in the same spirit one can use the above suggested fractional Mellin transform. Its advantage is that the integration variable, in the case of the Green's function $s\Rightarrow\xi_{ij}^0$, is relegated from its status as an exponent to the simple status of an ordinary variable in the denominator. This allows to write for the 4-point function, with the obvious transformations of variables $\zeta_{ij}=\log\mathbf{x}_{ij}^2$, the transformed argument of the original Green's function:  
\begin{equation}
G_{i_4\dots i_1}(\zeta_{i_1}\dots \zeta_{i_4})=\frac{1}{(2\pi i)^2}\int\mathrm{d}^2\xi M_{\kappa (i_4\dots i_1)}\big(\{\xi_{ij}\}\big)\prod_{i<j}\frac{\Gamma(\xi_{ij})}{(1+\zeta_{ij}\xi_{ij}^0/\kappa)^\kappa}
\end{equation}  
where the functions $\Gamma(\xi_{ij})$ are but appropriate normalisation factors. The integration is to be performed over the 2-dimensional surface of all the complex $\xi_{ij}$ with $
1\leq i<j\leq 4$ in this case, respecting the above kinematic condition on the total sum of the $\xi_{ij}$ and symmetries.  

Here we have used the same notation for the fractional Mellin amplitudes $M_{\kappa (i_4\dots i_1)}$  as in \citep{mack2009} and assumed that they are to be known from OPE, generalising them to the free parameter $\kappa$. These amplitudes are functionals of the functions $F$ and the kinematic relations on $\omega_i$. Determination of the fractional Mellin amplitudes is no trivial matter as it also is not in using the ordinary Mellin transform. At this stage one may return to the procedures of using the kinematic relations between the different exponents $\xi_{ij}^0$ and the identification of the Mellin amplitudes from field theory, methods that have meanwhile been described in extension in the literature. The calculation parallels that \citep[cf., e.g.,][]{fitz2012} for the ordinary Mellin analogue. It is clear that in taking the limit $\kappa\to\infty$ after having performed all calculations reproduces the result that would be obtained applying the ordinary Mellin transform. 

The possible advantage of the approach given in this brief note is that in the last expression the complex integration variables appear merely as variables in the denominator of the inverse fractional Mellin transform $\mathcal{M_\kappa}^{-1}$.  In addition to the freedom gained which is buried in the arbitrary, possibly even complex parameter $\kappa$, they have got rid of their property of being complex powers, a property that should substantially simplify the application to CFT and the CFT/AdS duality. The second form of the Mellin transform lacks this obvious advantage.


\begin{thebibliography}{99}
%
%
\bibitem[Treumann \& Baumjohann(2014)]{treumann2014} Treumann R A \& Baumjohann W. (2014) Fractional Laplace transforms -- a perspective, Front. Phys. 2:29. doi:10.3389/fphy.2014.00029

\bibitem[Mack(2009)]{mack2009} Mack G. (2009) D-independent representation of conformal field theories in D dimensions via transformation to auxiliary dual resonance models. Scalar amplitudes, arXiv:0907.2407v1 [hep-th]

\bibitem[Mack(2010)]{mack2010} Mack G. (2010) D-dimensional conformal field theories with anomalous dimensions as dual resonance models, arxiv:0909.1024v1 [hep-th]

\bibitem[Maldacena(1998)]{malda1998} Maldacena J M. (1998) The large N limit of superconformal field theories and supergravity, Adv. Theor. Mathemat. Phys. 2:231

\bibitem[Osterwalder \& Schrader(1973)]{oster1973} Osterwalder K \& Schrader R. (1973) Axioms for euclidean Green's functions 1. Comm. Math. Phys. 31:83-112

\bibitem[Osterwalder \& Schrader(1975)]{oster1975} Osterwalder K \& Schrader R. (1975) Axioms for euclidean Green's functions 2. Comm. Math. Phys. 42:281-305

\bibitem[Fitzpatrick \& Kapla(2012)]{fitz2012} Fitzpatrick, A L \& Kaplan J. (2012) Analyticity and the holographic S-matrix. J. High Energy Phys. 10:127, doi:10.1007/JHE P10(2012)127

\end{thebibliography}
\end{document}